\documentclass[%
 reprint,
 amsmath,amssymb,
 aps,
]{revtex4-1}

\usepackage{graphicx}
\usepackage{dcolumn}
\usepackage{bm}

\usepackage{blindtext}
\usepackage{amsmath}
\usepackage{amssymb}
\usepackage{physics}
\usepackage{siunitx}
\usepackage{subfig}
\usepackage{color}
\usepackage{booktabs}
\DeclareSIUnit\century{century}
\DeclareSIUnit\year{yr}
\newcommand{\keV}{\kilo\electronvolt}

\newcommand{\tit}[1]{\textit{#1}}

\newcommand{\pwo}{PbWO$_4$}

\newcommand{\sr}{$^{90}$Sr}
\newcommand{\co}{$^{60}$Co}
\newcommand{\eu}{$^{152}$Eu}
\newcommand{\cs}{$^{137}$Cs}
\newcommand{\ba}{$^{133}$Ba}
\newcommand{\na}{$^{22}$Na}
\newcommand{\ka}{$^{40}$Ka}

\begin{document}


\title{Development of a SiPM-based readout-module for the characterization of various scintillator materials}

\author{Lukas Nies}
\email{Lukas.Nies@physik.uni-giessen.de}
\author{Hans-Georg Zaunick}%
\author{Kai-Thomas Brinkmann}
\affiliation{%
 II. Institute of Physics, University of Giessen\\
}%

\date{\today}

\begin{abstract}

In this paper we discuss a general approach to use SiPMs in different electrical configurations in combination with numerous scintillator materials for applications in nuclear instrumentation, such as calorimetry and timing. The ``hybrid" configuration used in the PANDA detector was found to show the fastest timing properties with several hundred $\si{\pico\second}$ response by combining the advantages of a large active area while avoiding high operation voltages and dark noise. A ``parallel" configuration shows the largest light collection efficiency and is therefore suitable for energy measurements. Several calibration spectra were taken to measure the energy deposit of minimal ionizing cosmic muons in a $2$-$\si{\centi\meter}$-thick \pwo{} crystal.  \par

\end{abstract}

\maketitle

\begin{figure}[t!]
	\subfloat[Raw signal of the different configurations: blue (4x1 hybrid), red (4x1 parallel), green (1x1 single)] {\includegraphics[width=0.48\textwidth]{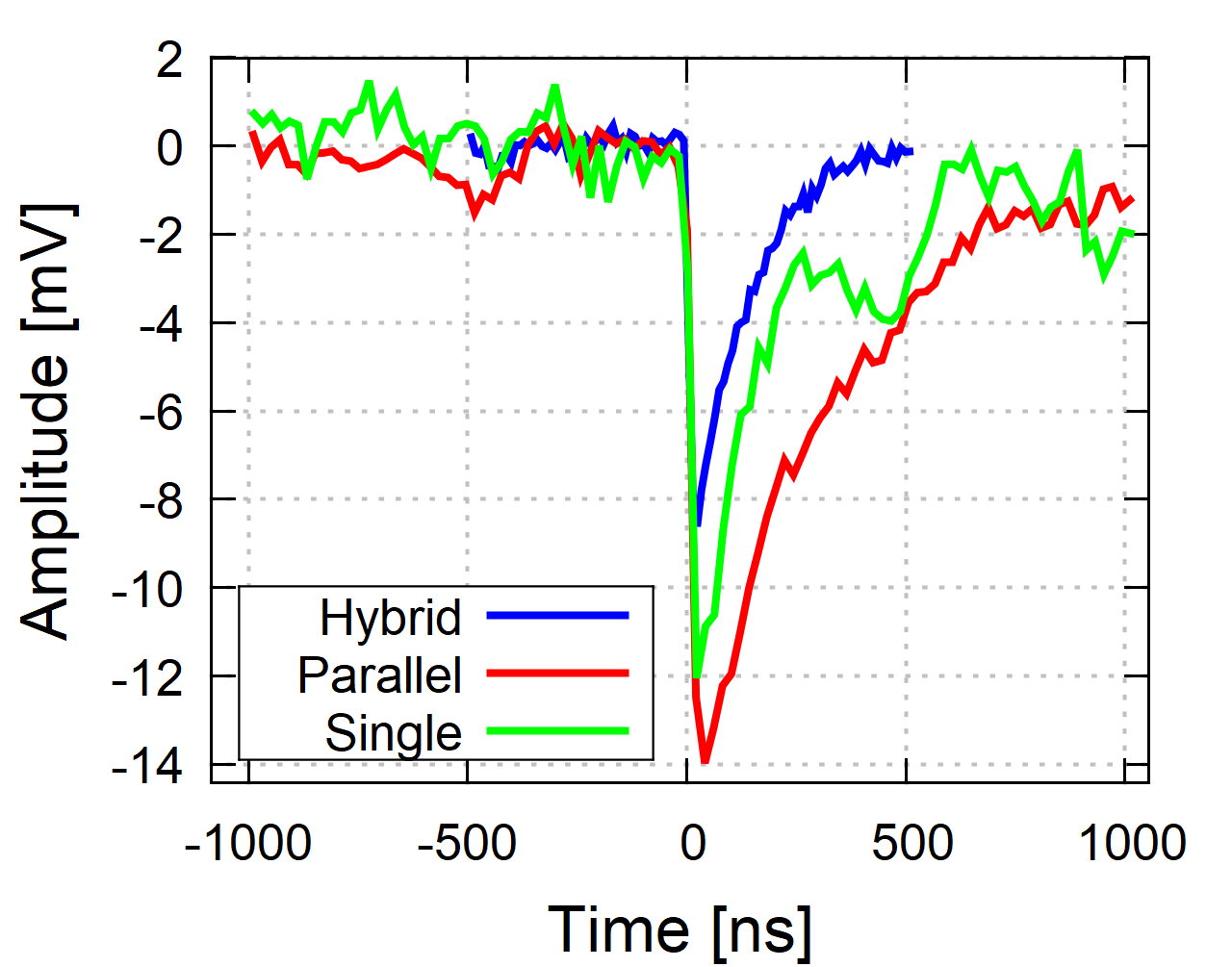}}
	\hfill
	\subfloat[Amplified and normalized signal of blue (4x1 hybrid), red (4x1 parallel), green (1x1 single) with utilized photonique preamplifier] {\includegraphics[width=0.48\textwidth]{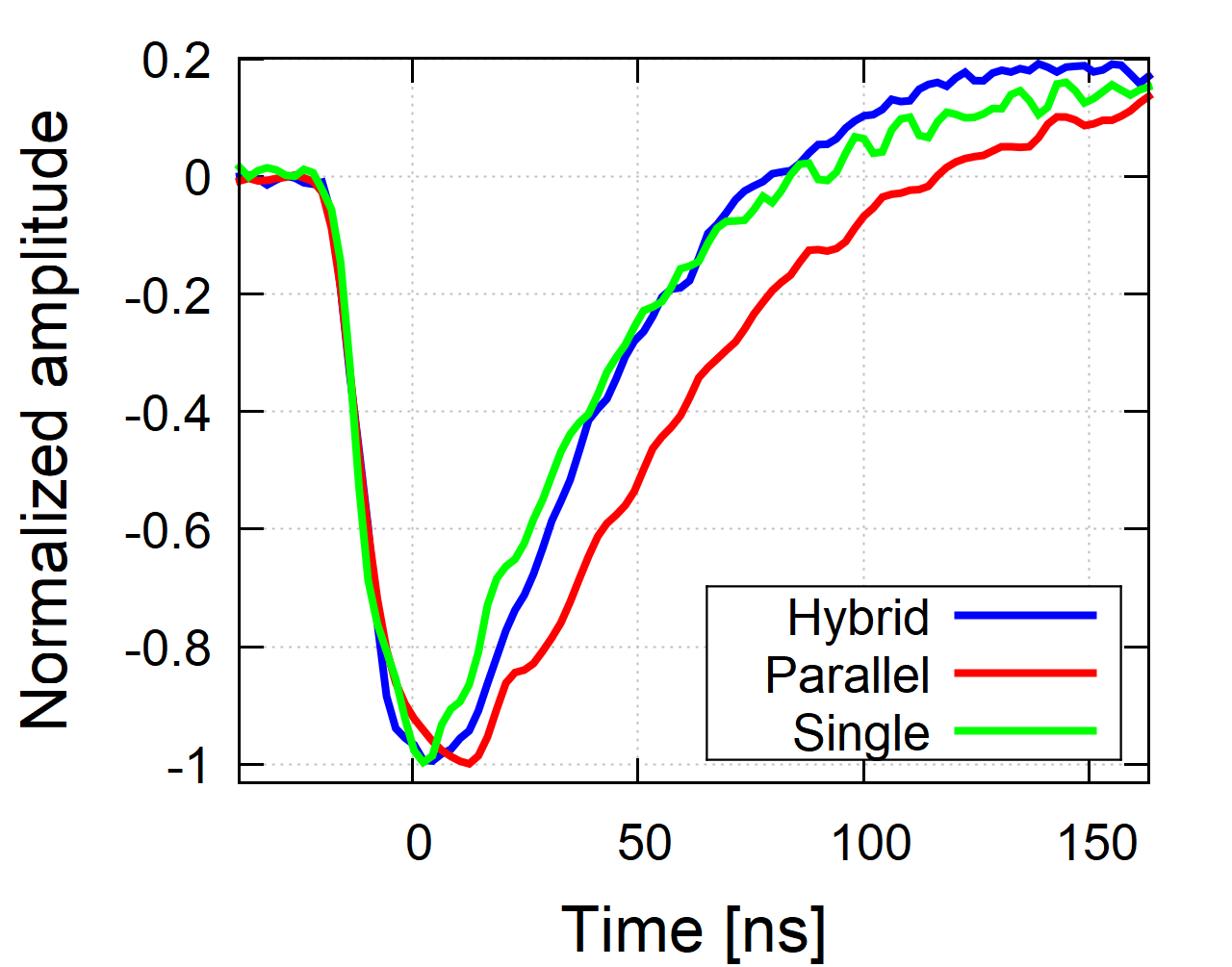}}
	\hfill
	\caption{Raw and amplified signals of different SiPM configurations.}
	\label{fig:signals}
\end{figure}

\begin{figure}[t!]
	\subfloat[Measurement positions] {\includegraphics[width=0.24\textwidth]{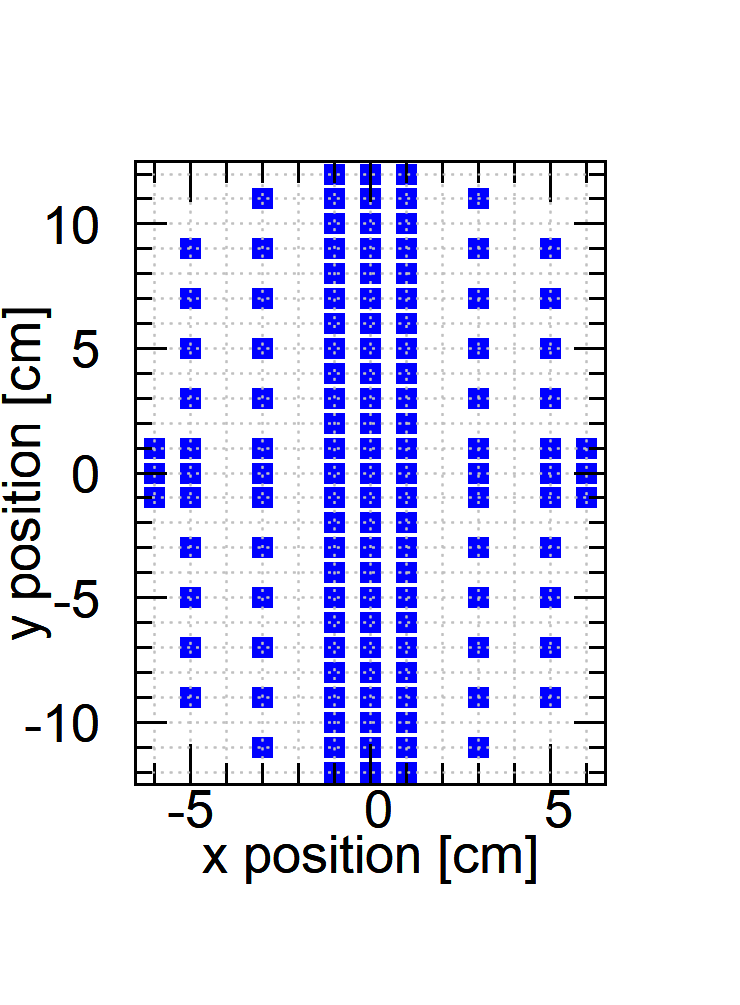}}
	\hfill
	\subfloat[Detector one, position at $y=+\SI{12}{\centi\meter}$] {\includegraphics[width=0.24\textwidth]{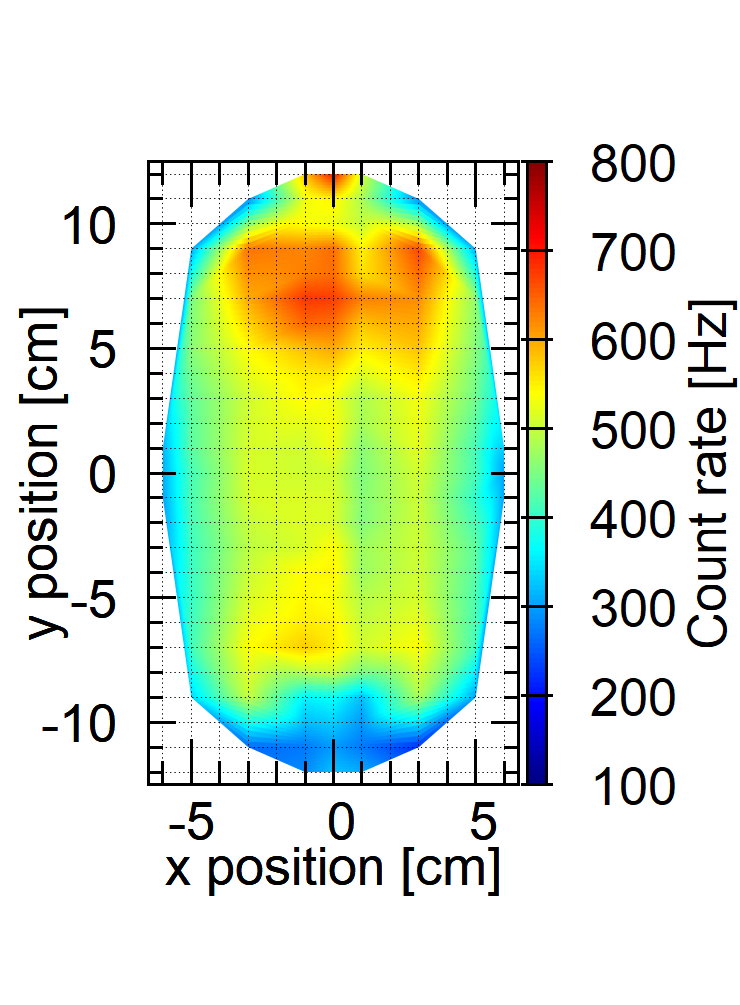}}
	\hfill
	\subfloat[Detector two, position at $y=-\SI{12}{\centi\meter}$] {\includegraphics[width=0.24\textwidth]{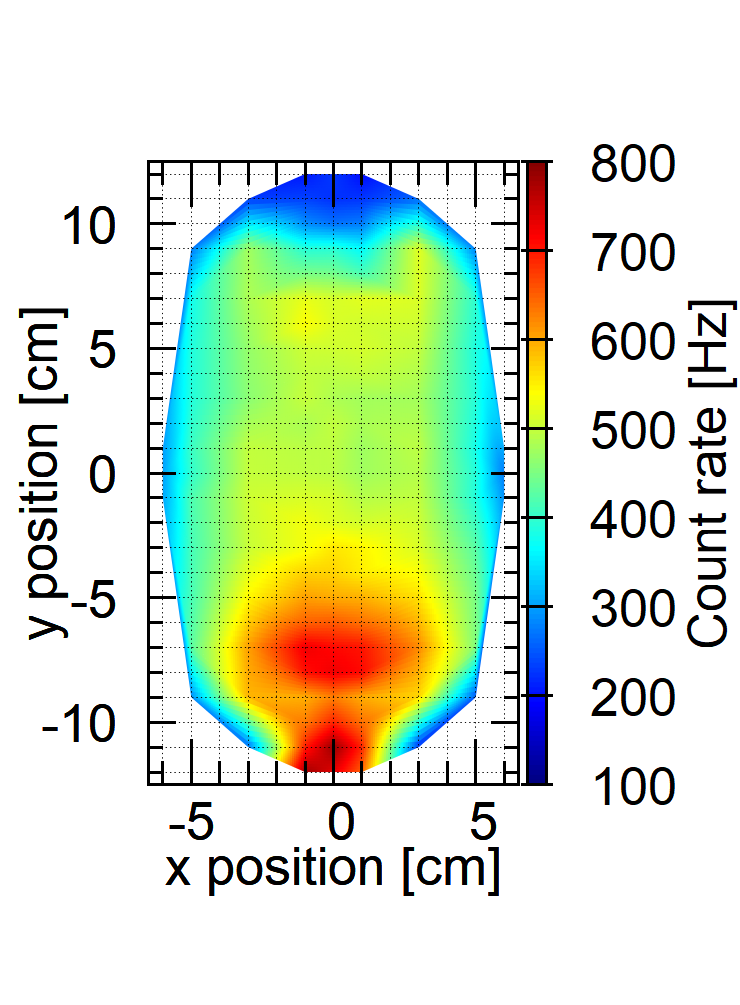}}
	\hfill
	\subfloat[Coincidence rate, same detector layout] {\includegraphics[width=0.24\textwidth]{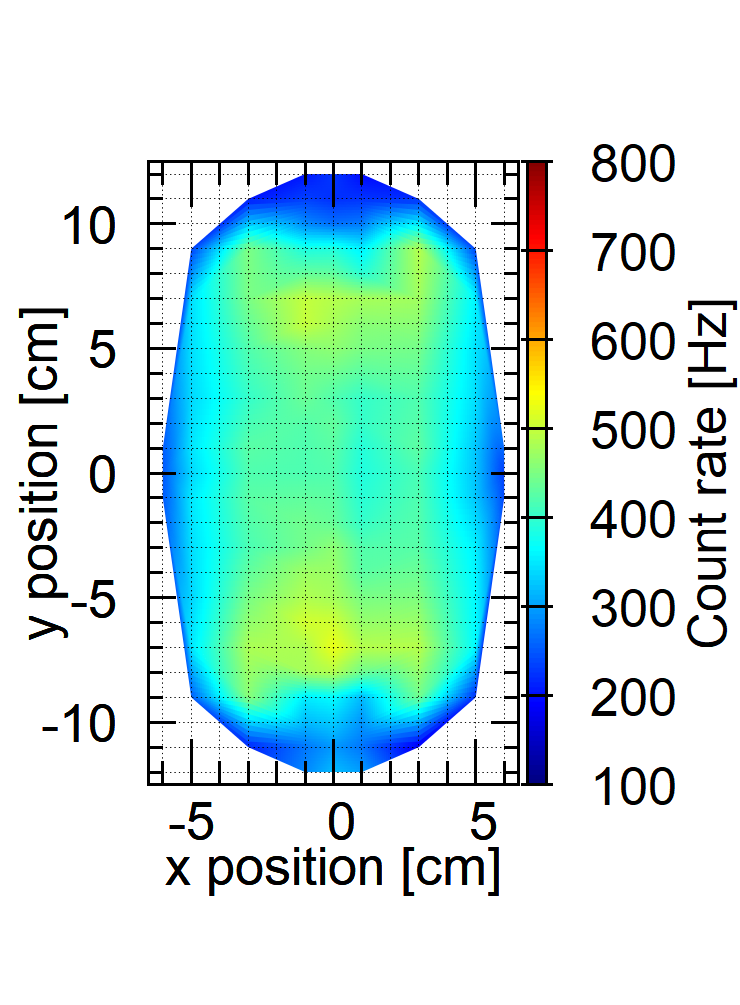}}
	\hfill
	\caption[Efficiency measurement setup]{Efficiency measurement at plastic scintillator with a \sr{} source at different positions. The heat map was plotted with \tit{gnuplot's pm3d} feature, the data between the measurement points was interpolated by the same software. Note that the coincidence count rate in front of one detector is mainly dominated by the count rate of the opposite detector.}
	\label{fig:efficiency}
\end{figure}

\begin{figure}[t!]
	\centering
	\subfloat[Measurement positions] {\includegraphics[width=0.24\textwidth]{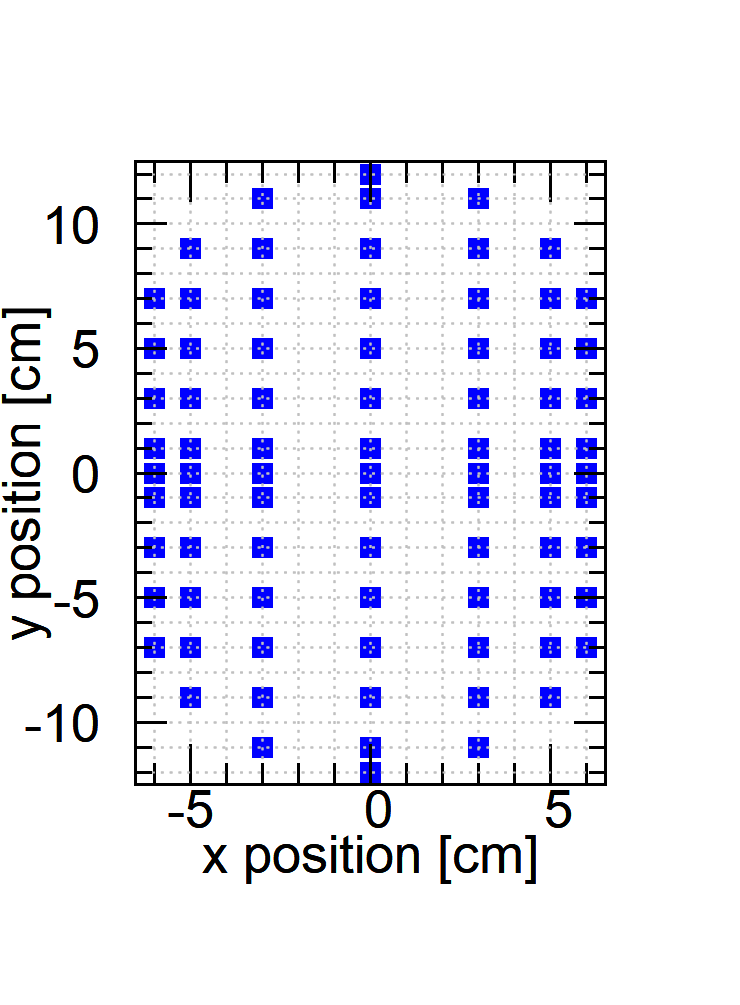}}
	\hfill
	\subfloat[Time resolution] {\includegraphics[width=0.24\textwidth]{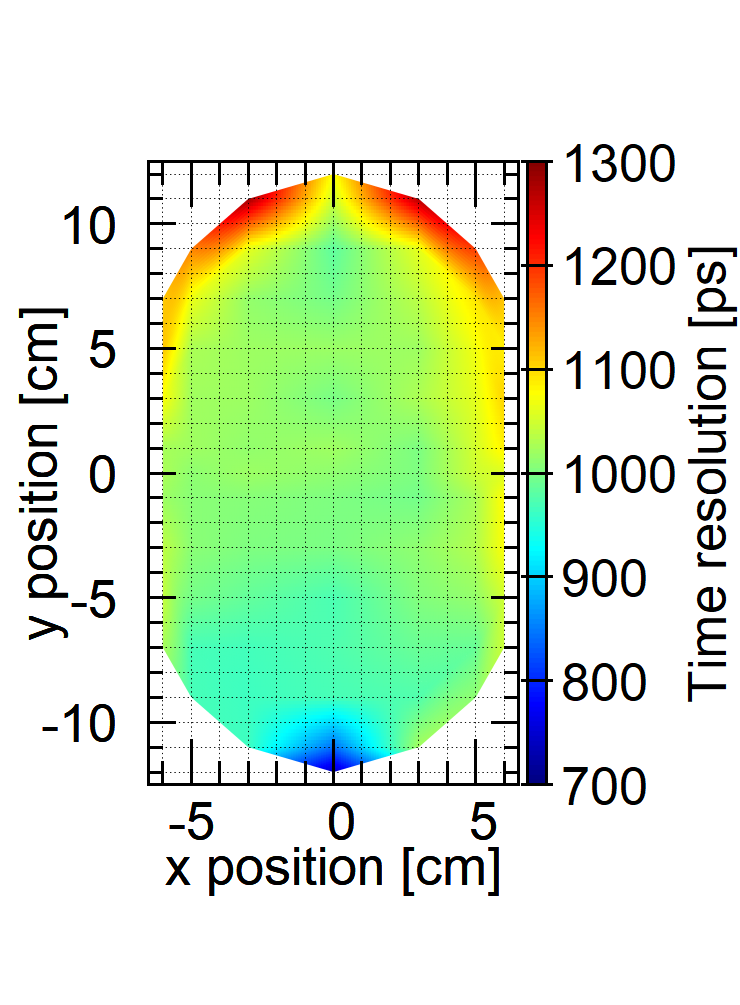}}
	\hfill
	\subfloat[Propagation time]
	{\includegraphics[width=0.24\textwidth]{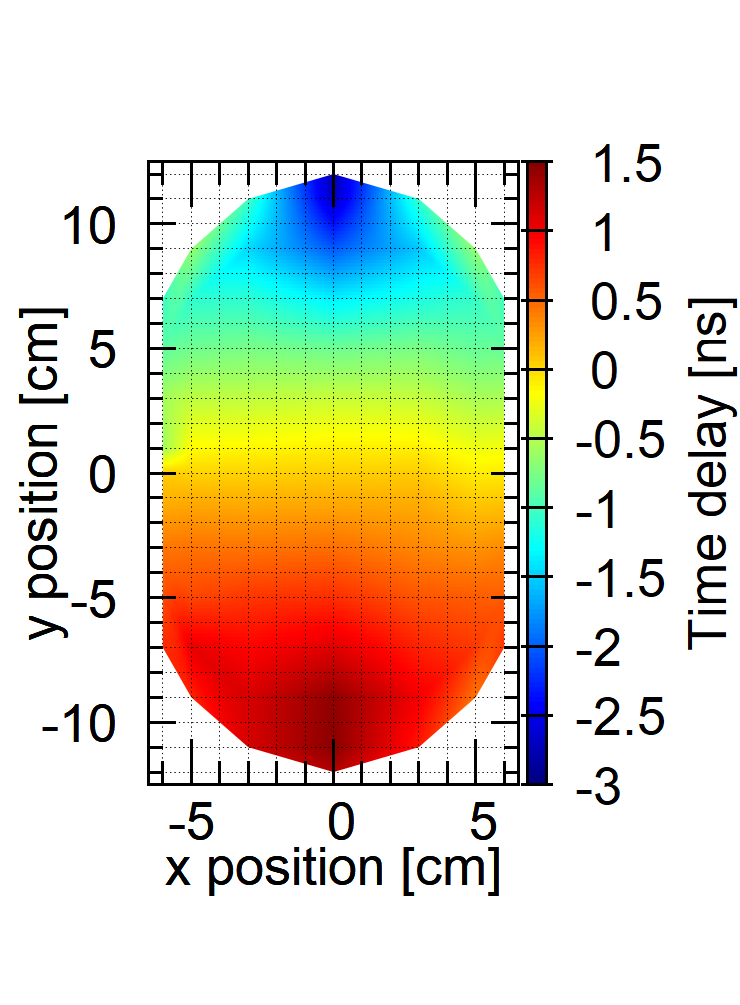}}
	\hfill
	\caption[Timing reolution measurement]{Time resolution measurement at plastic scintillator with a \sr{} source at different positions. The heat map was plotted with \tit{gnuplot's pm3d} feature, the data between the data points was interpolated by the same software. Detector board one is at position $y=\SI{+12}{\centi\meter}$ and board two at $y=\SI{-12}{\centi\meter}$}
	\label{fig:timing}
\end{figure}

\begin{figure}[b!]
	\centering
	\includegraphics[width=0.9\linewidth]{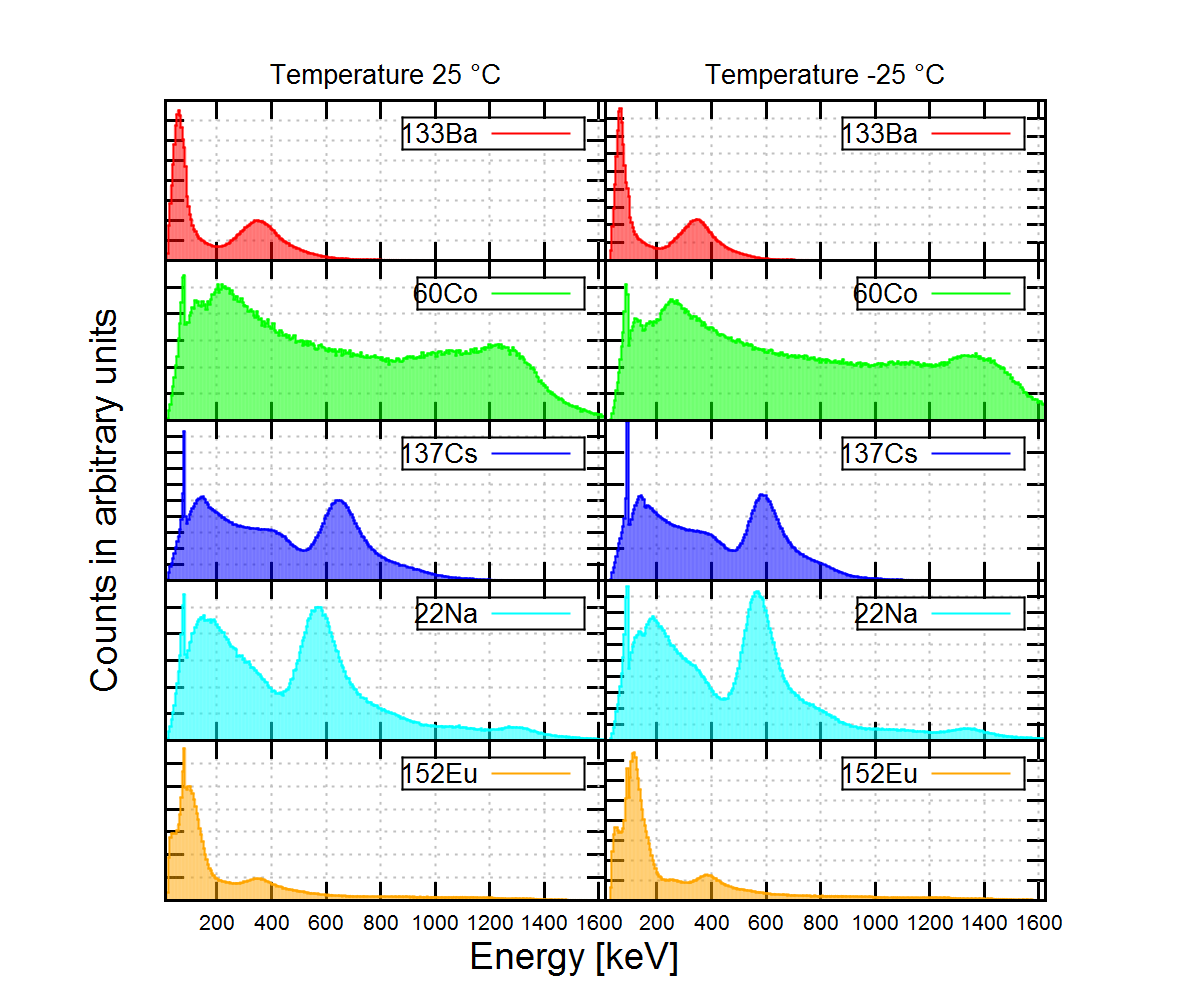}
	\caption{Energy spectra vs temperature of several calibration sources using the LYSO crystal.}
	\label{fig:lyso_energy}
\end{figure}

Modern photodetector applications encounter a large spectrum of different experimental environments. Prerequisites, e.g. working within strong magnetic fields, providing large intrinsic amplification, independence on high operation voltage, and space restrictions led to the development of a new type of detector, the semiconductor diodes. The \tit{Silicon Photomultiplier} (SiPM) integrates a large number of \tit{avalanche photodiodes} as microcells within a small space. With a large intrinsic amplification of up to $10^6$, the SiPM comes in many different packaging sizes and works mainly with low voltage between $\SI{25}{\volt}$ and $\SI{85}{\volt}$. Its insensitivity to magnetic fields and single photon counting capability in combination with suited scintillators make SiPMs a valuable choice for modern challenging applications, i.e. nuclear magnetic resonance imaging, positron emission tomography, and high energy calorimetry and tracking. \\ \indent
The development of the scintillation tile hodoscope \cite{SciTil}, short SciTil, for the \tit{PANDA} detector at the future Facility for Antiproton and Ion Research (FAIR) at the GSI Helmholtz Center for Heavy Ion Research in Germany \cite{FAIR} necessitated the invention of a small-scale photo readout system insensitive to the magnetic field of the detector. Silicon photomultiplier were chosen due to previously mentioned properties. Coupling multiple SiPMs in an array increases the effective active area and enhances the photon detection efficiency. \\ \indent
\begin{figure}[b!]
	\centering
	\includegraphics[width=0.5\linewidth]{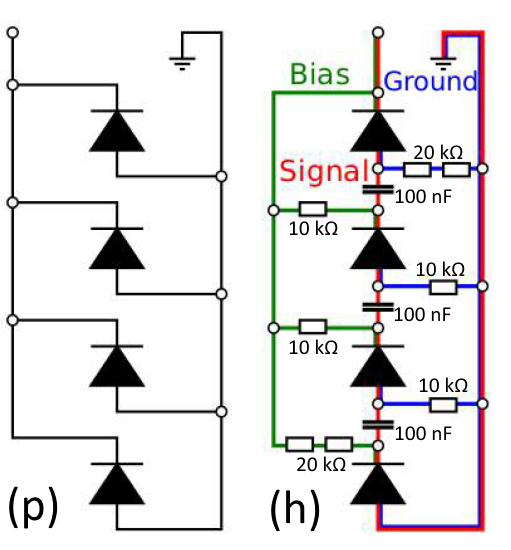}
	\caption{ Schematic of the two different SiPM configurations \cite{sebastian}. On the left is the ``parallel" configuration, on the right is the ``hybrid" configuration. The signal path is depicted in red, the bias current in green, and the ground in blue.  }
	\label{fig:PCB}
\end{figure}
In order to develop a general purpose SiPM array we designed in first iteration a printed circuit board (PCB) which is able to host either a ``parallel" or ``hybrid" configuration by just adding or removing some resistors and capacitors on the back of the board (see figure \ref{fig:PCB}). In parallel configuration the output signal is the sum of the output of all diodes. A faster signal is received in hybrid configuration when capacitors connect anode and cathode of adjacent diodes and resistors are used for connecting the SiPMs with bias and ground. The signal is read out via a ``bias-T" consisting of a resistor and a capacitor which decouple the signal from the DC biasing. In this paper, KETEK PM3350-EB diodes with an active area of $\SI{3}{\milli\meter}\times\SI{3}{\milli\meter}$, a pixel size of $\SI{50}{\micro\meter}\times\SI{50}{\micro\meter}$, and 3472 pixels were used \cite{SiPM_Manual}. \\ \indent
Attached to a plastic scintillator, comparing the raw and normalized signal shapes of a single SiPM with an array of four diodes in either parallel or hybrid configuration shows the general differences of the methods. The signal amplitude for the hybrid configuration is roughly $\frac{1}{\sqrt{N}}$ compared to the amplitude of a single SiPM, where $N$ stands for the number of SiPMs per array. Driving the SiPM in parallel mode does not affect the signal amplitude. With a few nanoseconds the hybrid configuration has a considerably faster rising edge with respect to the parallel configuration, where the latter tends to get slower as the number of SiPMs per array increases. Hence, SiPM-arrays in hybrid configuration are more suitable for time measurements whereas the parallel configuration is useful for energy measurements. \\ \indent
In order to yield breakdown and operation voltages, the temperature dependent current-voltage characteristics were studied. For this, the different configurations were wrapped with opaque tape to prevent occurrence of photon induced currents and put in a programmable refrigerator. For different temperatures between $-25^{\circ}C $ and $25^{\circ}C$ the SiPM-boards were biased and the IV-curves were measured with a custom made high voltage distribution board designed for operating the avalanche photo diodes of the electromagnetic calorimeter barrel of the PANDA detector \cite{chris}. \\ \indent
If a photodiode is driven in reversed bias mode the depletion layer between the p- and n-junction broadens and the built-in potential intensifies. At the breakdown voltage $U_{BD}$ electrons gain sufficient energy to create a self-sustained avalanche which leads to an exponential increase in current flowing. The breakdown voltages were determined by the intersection of a firstorder polynomial fit for the voltage area before breakdown and a secondorder polynomial for the region right after breakdown. \\ \indent
In order to find the optimal operation voltage $U_{OP}$, which is usually a few volts beyond breakdown, the minimum of the relative slope $\frac{dI}{dV}\frac{1}{I(V)}$ of the IV-curve was fitted. \\ \indent 
After scanning the IV-characteristics at different temperatures, the resulting breakdown temperature coefficient $k_{BD}$ for a single SiPM was in good agreement with the specifications of the manufacturer \cite{SiPM_Manual} (see \ref{tab:IV_cata}). The value for multiple SiPMs does not seem to change (within the error). The coefficient for the operation voltage seems to add as the number of SiPMs increases: the four-SiPM array has a roughly four times higher value as the single SiPM. \\ \indent
\begin{table}[t!]
	\centering
	\begin{tabular}{ c|cc|cc } \toprule[2pt]
		& \multicolumn{2}{c|}{(s)-configuration} & \multicolumn{2}{c}{(h)- and (p)-configuration} \\
		T [$\si{\degreeCelsius}]$ & $V_{OP}$ [$\si{\volt}$] & $V_{BD}$ [$\si{\volt}$] & $V_{OP}$ [$\si{\volt}$] & $V_{BD}$ [$\si{\volt}$]  \\ \midrule
		-25 & $30.78\pm 3.99$ & $24.62\pm 2.89$ & $30.72\pm 0.47$ & $24.40\pm 3.67$ \\
		0 & $30.67\pm 0.42$ & $25.50\pm 0.60$ & $30.89\pm 0.15$ & $24.78\pm 0.03$ \\
		5 & $30.72\pm 0.32$ & $24.93\pm 0.15$ & $31.01\pm 0.59$ & $24.87\pm 0.04$ \\
		10 & $30.70\pm 0.35$ & $25.11\pm 0.44$ & $31.20\pm 0.32$ & $25.04\pm 0.04$ \\
		15 & $30.74\pm 0.28$ & $25.30\pm 0.14$ & $31.32\pm 0.68$ & $25.08\pm 0.01$ \\
		20 & $30.80\pm 0.23$ & $25.35\pm 0.05$ & $31.57\pm 0.75$ & $25.18\pm 0.03$ \\
		25 & $30.87\pm 0.45$ & $25.51\pm 0.23$ & $31.04\pm 2.11$ & $25.20\pm 0.06$ \\
		\midrule
		& $k_{OP}$ [$\si{\milli\volt\per\kelvin}$] & $k_{BD}$ [$\si{\milli\volt\per\kelvin}$] & $k_{OP}$ [$\si{\milli\volt\per\kelvin}$] & $k_{BD}$ [$\si{\milli\volt\per\kelvin}$]  \\
		& $6.79\pm 1.34$ & $23.66\pm 5.10$ & $23.67\pm 2.90$ & $19.30\pm 1.51$ \\
		\bottomrule[2pt]
	\end{tabular}
	\caption[Fit results for operation and breakdown voltages]{Fit results for operation and breakdown voltages. Since the same SiPMs were used for (h)- and (p)-configuration the results are similar.}
	\label{tab:IV_cata}
\end{table}  
Since the diode is driven a few volts beyond breakdown a probability is given that an avalanche is triggered by high energetic seed electrons (free electrons in the silicon with sufficient thermal energy). Because a SiPM is made up of thousands of microcells it is very likely to have plenty of breakdowns in parallel every second. This effect can be suppressed by cooling where the high energetic part of the Boltzmann distributed electrons in the depletion layer gets shifted to lower thermal energies. \\ \indent
\begin{figure}[b!]
	\centering
	\includegraphics[width=0.9\linewidth]{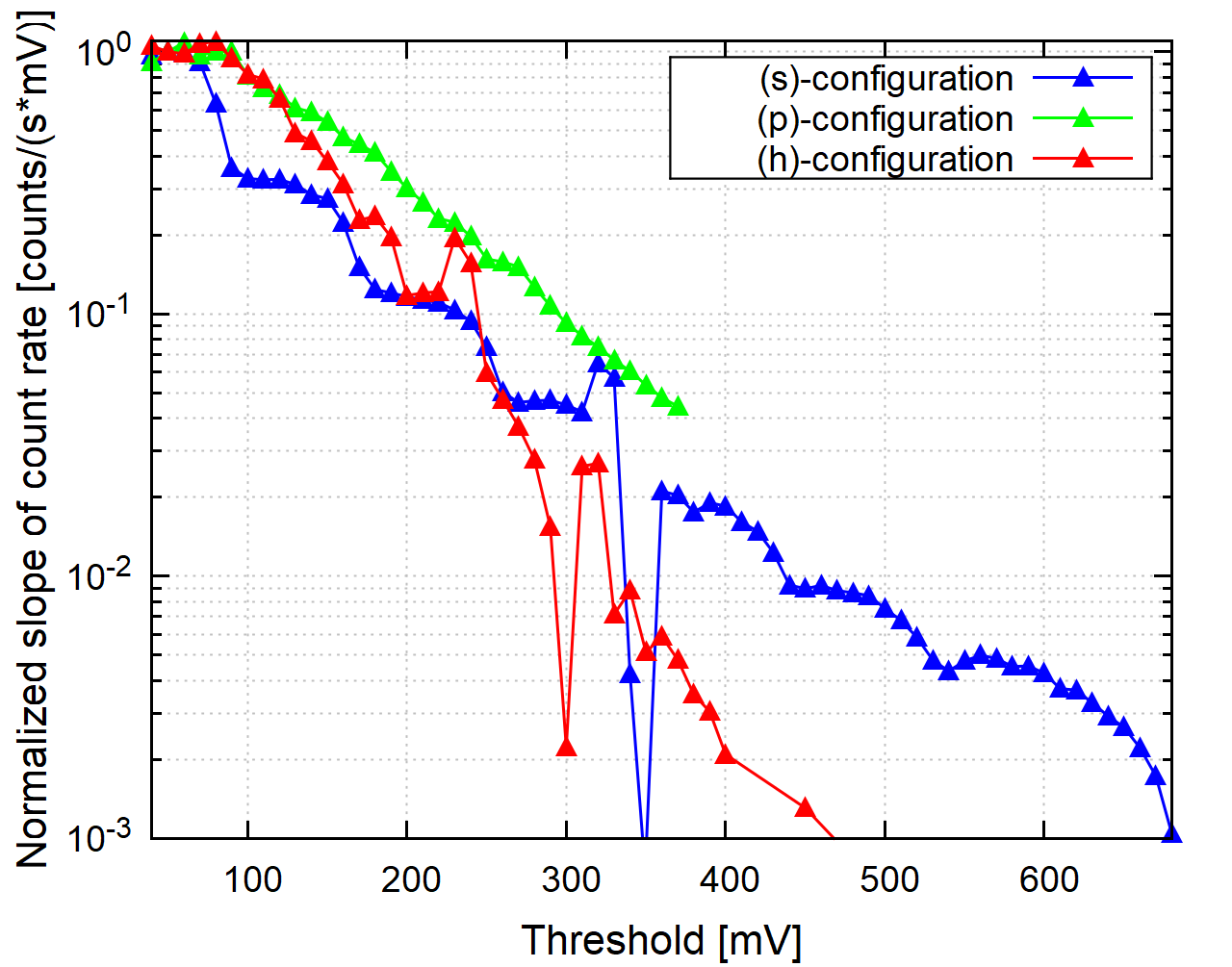}
	\caption{Normalized slope of the dark count rate (note the logarithmic scale). The distinct step-like behavior for a single SiPM (blue) is clearly visible, while this subtle effect vanishes for multiple SiPMs (hybrid in red, parallel in green).}
	\label{fig:dark_count_rate}
\end{figure} 
Examining this so-called dark count rate (signal without seeing a photon) of SiPMs is crucial for light sensitive applications. It is not recommended using SiPMs at room temperature for single photon counting since the dark count rate is high. By deploying different diode configurations in an opaque box and measuring the pulse height spectrum behind a pre-amplifier with an oscillator in persistence mode we investigated the dependence of count rate and threshold. For better visibility of this effect, the normalized dark count rate for different configurations is depicted in figure \ref{fig:dark_count_rate}. Because every microcell, if firing, contributes the same signal to the overall signal a single SiPM shows a step-like behavior for increasing thresholds. The possibility of three cells firing at the same time is suppressed by a factor of ten, and for six cells the suppression is almost a factor of a hundred. In our setup, this effect cannot be seen for multiple SiPMs. Every diode is biased by the same voltage, but each SiPM has slightly different gain. Therefore the pulse heights smear and a threshold scan does not yield a step-like drop in count rate. \\ \indent
For precise measurements with SiPMs at scintillators with low expected light yield, it is necessary to match diodes per array with comparable gain and to cool the detector for lower noise. \\ \indent
In order to examine the timing properties of the configurations, a plastic scintillator EJ-248M from ELJEN with sizes $\SI{25.50}{\centi\meter}\times\SI{12.25}{\centi\meter}\times\SI{5}{\milli\meter}$ was cut (for geometry see figure \ref{fig:timing}), wrapped in layers of Teflon, aluminum, and finally in thick, opaque foil. Two opposite sides, which has been designed to have the same area as the SiPM-boards, were left uncovered. For spacing, the boards were equipped with a rubber mask with reflective foil leaving the SiPMs uncovered. For detailed descriptions see \cite{Lukas_Thesis}. This design was chosen for measuring cosmic muons for the project \tit{Cosmic Radiation: Measuring Cosmic Muons with the Raspberry Pi} \cite{schauer_projekt}, because this shape provides a large area with a ``pseudo" light-guide towards the narrow edges where the SiPM boards were attached. Optical grease was used to ensure good light transmission between scintillator material and SiPM entrance window. \\ \indent 
First, the raw signals of the boards were examined. A "bias-T" \cite{Lukas_Thesis} decouples the AC signal from the DC biasing. Afterwards, pre-amplifier from \tit{Photonique} \cite{photonique} were used for signal shaping. The results can be seen in figure \ref{fig:signals}. \\ \indent
As mentioned above, the signal strength depends on the number of SiPMs per board and on the configuration. The amplitude for the hybrid configuration gets divided by the square root of number of SiPMs utilized in the circuit where the parallel configuration has the same strength as a single SiPM. When using a preamplifier and normalizing the spectra one can see that the rise time of the hybrid board is fastest and the parallel board is slowest. Again, this suggests the use of hybrid boards for timing. \\ \indent
In preparation for the timing measurement the symmetry and the counting efficiency was tested. For this, two hybrid boards were mounted on each side of the scintillator and were connected via pre-amplifiers to a leading edge discriminator. The outputs of those two channels were fed into a counter, which was controlled by an external clock. A third channel was used to count the coincident counts of channel one and two. The threshold of the leading edge discriminator was set for both channels to account for the expected rate of cosmic muons in the scintillator of approximately ($\SI{200}{\frac{1}{\second\meter\squared}}$ \cite{PDG}) . A \sr{} source with a small collimator was (manually) placed on several areas on top of the scintillator to create scintillation light (see figure \ref{fig:efficiency}(a)). The count rate was determined by measuring $\SI{10}{\second}$ per position.\\ \indent
The heat map plots in figure \ref{fig:efficiency} for the count rate show two maxima for each detector: one right in front of the SiPMs, the other on the y-axis some centimeters in front of the detectors, beginning half the way down from the center of the scintillator to the mounted boards. This region of high count rate is limited to some square centimeters and is oval shaped. The count rate in the other regions of the plastic is nearly constant, except from the very edges and corners, and right in front of the respective opposite detector, where the rate drops. The results for detector one and two are similar but not identical: detector one seems to be less effective in counting compared to detector two, which shows a more symmetric shape. The second maximum might be explained by the special shape of the cut corners, so the photons generated at this position might get reflected towards the detectors. \\ \indent 
In discussing the coincident count rate, the asymmetry can be found again: the rate right in front of detector two is nearly twice as high as in front of detector one. The maxima appear again at the same positions, the rate is highest on the y-axis and decreases for positions moving along the x-axis. For events right in front of one board, the coincident count rate is apparently given by the count rate of the opposite board. \\ \indent  
Although one of the two detectors seems to have some performance issues, the coincident count rate for most of the positions is a square from $\SI{-5}{\centi\meter}$ to $\SI{+5}{\centi\meter}$ on the x-axis and $\SI{-9}{\centi\meter}$ to $\SI{+9}{\centi\meter}$ on the y-axis is valid for using this setup for measuring cosmic muons (two detectors are used to avert detecting fake/noise events). \\ \indent
Some applications, e.g. position and timing measurements, necessitate sufficient time resolution in the range of picoseconds. To test time resolution properties of the plastic scintillator setup the readout electronics had to be slightly changed. The leading edge discriminator (LED) was exchanged by a constant fraction discriminator (CFD) to avoid walk. The threshold was chosen to be slightly above noise to detect the very first arriving photons. The outputs of the two channels were fed into time to pulse height converters (TPC) where channel one served as the start and channel two, after a cable delay of fixed length, as the stop signal. The coincidence signal of both channels was used as a gate to avoid noise. The output of the converter was digitized by an analog to digital converter (ADC) and processed in a \tit{root} environment on a personal computer. To calibrate the net delay between channel one and delayed channel two, a \sr{} source was placed in the middle of the detector. Testing different delays, the zero net delay was set to lie in the center of the acceptance range of the 13-bit ADC. \\ \indent
To test the time resolution the \sr{} source was move along the y-axis across the detector and the time difference between detector one and two was measured. The resolution was extracted from the standard deviation of a Gaussian fit to the time pulse spectrum of the TPC output.   
The results are plotted in figure \ref{fig:timing}(b). Again, asymmetrical behavior is found: the resolution is roughly constant ($\SI{1}{\nano\second}$) in most parts of the detector and better towards the lower detector ($\SI{0.7}{\nano\second}$) but worse in regions around the upper detector. \\ \indent
Furthermore, the propagation time of scintillation light within the scintillator plate was examined. Calculations give a maximal travel time of roughly $T_\text{max}=\SI{1300}{\pico\second}$ \cite{Lukas_Thesis} from one detector board to the other, which corresponds to a distance of $\SI{25}{\centi\meter}$. Placing the source in the geometrical center results in a zero time delay (see figure \ref{fig:timing}(c)). Approaching a board increases the delay until the maximal position above one board is reached where the time delay should be $T_{\text{max}}$. This can be observed for board one but not for board two where the time difference is close to $2\times T_{\text{max}}$.\\ \indent 
An explanation for the asymmetry in the efficiency, time resolution and propagation time measurements might be dissimilar sensitivities of the two detector boards. A good resolution is achieved when both boards are exposed to a high amount of direct light, so seeing less direct light results in bad timing. One can assume, and the efficiency measurements confirms this, that, when the source is placed right in front of one SiPM-board, this set of diodes sees sufficient direct light. Hence, bad timing in front of one board is due to bad performance of the opposite board. Therefore, diode one seems to perform better than its counterpart. \\ \indent
This problem might be overcome by fine tuning the thresholds and gains (biasing) of the detectors and matching the SiPMs per array. Further work on tuning the setup parameters as well as using up to date processing electronics might lead to a distinct increase in precision. Also, using a light guide or embedded scintillating fibers could improve overall performance.\\ \indent
Another application for SiPMs could be usage in calorimetry. Examining the feasibility in this field, two prominent organic scintillator, LYSO ($\SI{10}{\milli\meter}\times\SI{10}{\milli\meter}\times\SI{3.5}{\milli\meter}$) and \pwo{} ($\SI{20}{\milli\meter}\times\SI{20}{\milli\meter}\times\SI{50}{\milli\meter}$), were equipped with SiPM-boards in parallel configuration ($2\times 1$ for LYSO and $3\times 2$ for \pwo{}) to maximize light collection. The boards were coupled to the crystals with optical grease and the whole ensemble was wrapped in teflon and opaque black tape. \\ \indent
The crystals were set up in the climate chamber for temperature stabilization. The signal, amplified by preamplifiers, was further shaped by a linear amplifier with adjustable integration time and gain. The energy spectrum was taken by an analog-to-digital converter from CAEN (N957) which was controlled by a UNIX-workstation. \\ \indent
First, the LYSO crystal was tested at $\SI{-25}{\degreeCelsius}$ under $\gamma$-radiation from a \co{} source. It appeared that the raw signal of the two SiPMs in parallel was so strong that the preamplifier saturated, so the bias voltage was set to $\SI{26}{\volt}$ right beyond breakdown. The same problem with the linear amplifier was solved by using a $\SI{-20}{\decibel}$ passive attenuator. \\ \indent  
Consecutively, the energy spectra of several calibration sources (\co{}, \na{}, \cs{}, \ba{} and \eu{}) were taken. To calibrate the setup, the most prominent photopeaks were fitted with Gaussians. The spectra are plotted in figure \ref{fig:lyso_energy}. The same procedure was repeated at $\SI{25}{\degreeCelsius}$. \\ \indent
The photopeak of \ba{} is distinct as well as the x-ray peak at low energies. The cobalt spectrum has a large compton background since the gamma quanta with $\SI{1.1}{\MeV}$ and $\SI{1.3}{\MeV}$ (all gamma ray energy data is retrieved from \cite{gamma_energy}) are too energetic to be fully stopped in the $\SI{3.5}{\milli\meter}$ thick crystal. For this, the two photopeaks are barely visible, the sum peak could not be observed at all. Furthermore, \cs{} and \na{} show a clear photopeak and a distinct compton spectrum, the $\SI{511}{\keV}$ peak due to annihilation of the positron from the $\beta^+$ decay of sodium is clearly visible. The \eu{} source was tested as a calibration source since the spectrum shows multiple gamma energies and it was found that the most prominent peaks ($\SI{344}{\keV}$, $\SI{244}{\keV}$, $\SI{121}{\keV}$, $\SI{40}{\keV}$) could be resolved, especially for $\SI{-25}{\degreeCelsius}$. The resolution of the cooled system is in some cases ($\SI{300}{\keV}$ to $\SI{700}{\keV}$) slightly better. For low and high energies, larger errors can be found. \\ \indent
Cooling does not seem to improve the energy resolution significantly, despite the intrinsic dark noise of the SiPM is suppressed and the light yield of the scintillator is improved. The strong raw signal indicates that a lot of scintillation light is seen by the SiPMs. Due to the high light output the SiPMs seem to be overexerted due to this large amount of photons. This might be solved by a higher number of microcells (smaller pixel size), so more photons can be detected at the same time. This overextension could be the reason why cooling does not result in a significant better resolution. Still, an improvement by cooling is expected for crystals with lower light yield, i.e. lead tungstate. \\ \indent
The lead tungstate setup with six parallel SiPMs is operated comparable to the LYSO setup with two SiPMs. However, the light yield of \pwo{} is several orders of magnitude lower than LYSO, and roughly $85-110$ photons per $\si{\MeV}$ are expected \cite{panda_emctdr}. \\ \indent
\begin{figure}[b!]
	\centering
	\includegraphics[width=0.9\linewidth]{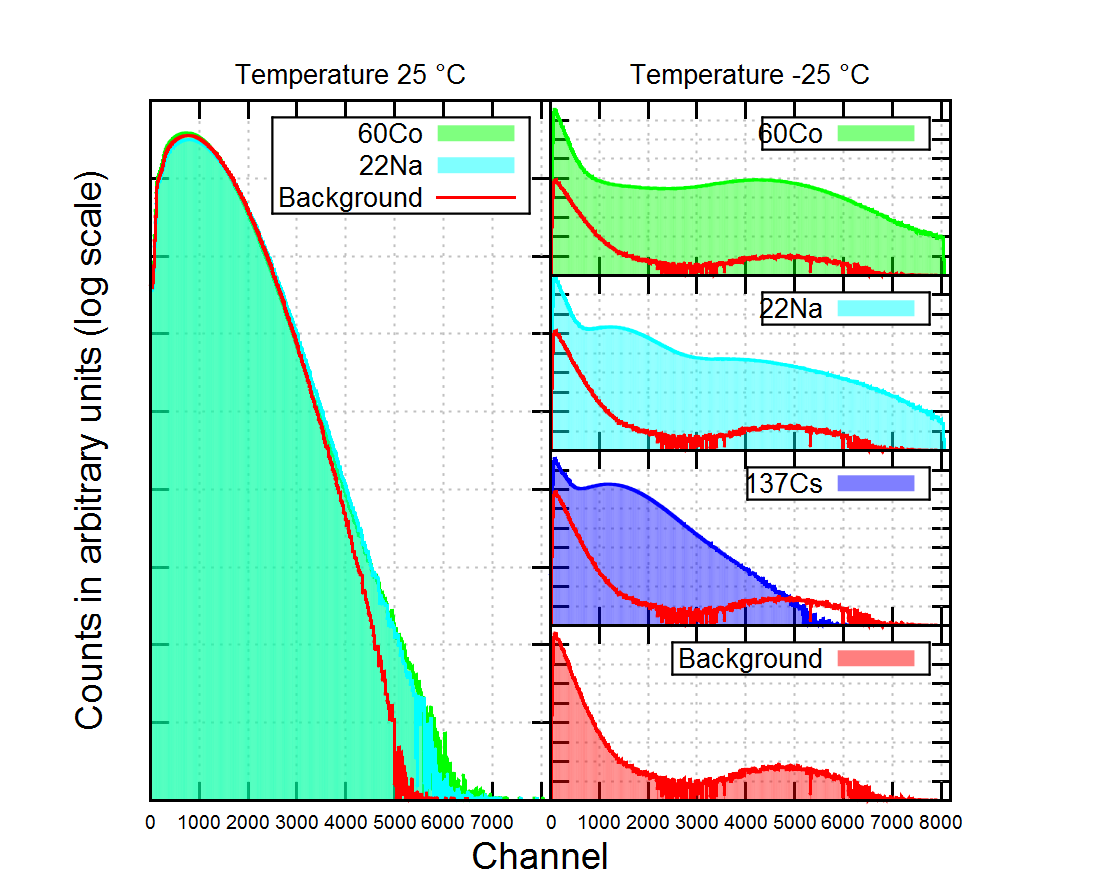}
	\caption[\pwo{} energy spectra]{\pwo{} energy spectra of several calibration sources.}
	\label{fig:ch5:pwo_energy}
\end{figure}
Figure \ref{fig:ch5:pwo_energy} shows the energy spectra of several calibration sources. At higher temperature ($\SI{25}{\degreeCelsius}$), the energy spectra merge with the background. At this point, one has to consider the \tit{photo detection efficiency} (PDE) of the SiPMs ($\approx 40\%$, dependent on bias voltage) for the emission wavelength of \pwo{} ($\SI{425}{\nano\meter}$), the loss of photons due to absorption ($\approx 10\%$) and the effective light sensitive area (ratio of the sensitive area of the SiPM and the $\SI{20}{\milli\meter}\times\SI{20}{\milli\meter}$ crystal surface, about $13.5$\%). Hence, only few photons are left, diminishing the resolution of the spectrum. The measurements at $\SI{25}{\degreeCelsius}$ show that only for a small amount of events a sufficient number of photons are registered to fill bins beyond the background. This is not sufficient for determining the particle energy. \\ \indent 
The spectra at $\SI{-25}{\degreeCelsius}$ are more distinct: \co{} shows a broad and flat peak in which the two photopeaks are merged. The \na{} spectrum shows both the annihilation peak and the photopeak at $\SI{1275}{\keV}$, so does the \cs{} spectrum at $\SI{662}{\keV}$. The background shows less noise compared to the background at $\SI{25}{\degreeCelsius}$. Reason being the higher light yield ($\approx \times 3.5$ \cite{panda_emctdr}) and the suppressed thermal noise of the SiPMs. A small peak at channel 5000 is also visible and will be discussed in the following. \\ \indent
The \pwo{} setup was also used to measure the energy deposit of cosmic muons in lead tungstate. According to \cite{panda_emctdr}, the energy deposit for minimum ionizing particles is $\SI{10.2}{\MeV\per\centi\meter}$, so one expects a mean energy loss of $\approx \SI{20}{\MeV}$ in the crystal with a thickness of $\SI{2}{\centi\meter}$. Since the energy deposit of cosmic muons in matter varies due to the statistical nature of the energy loss, whose mean value is given by the Bethe-Bloch-formula, a Landau distribution with a peak at $\approx \SI{20}{\MeV}$ is expected. \par
\begin{figure}[b!]
	\centering
	\subfloat[Energy spectrum without any trigger and small range] {\includegraphics[width=0.48\textwidth]{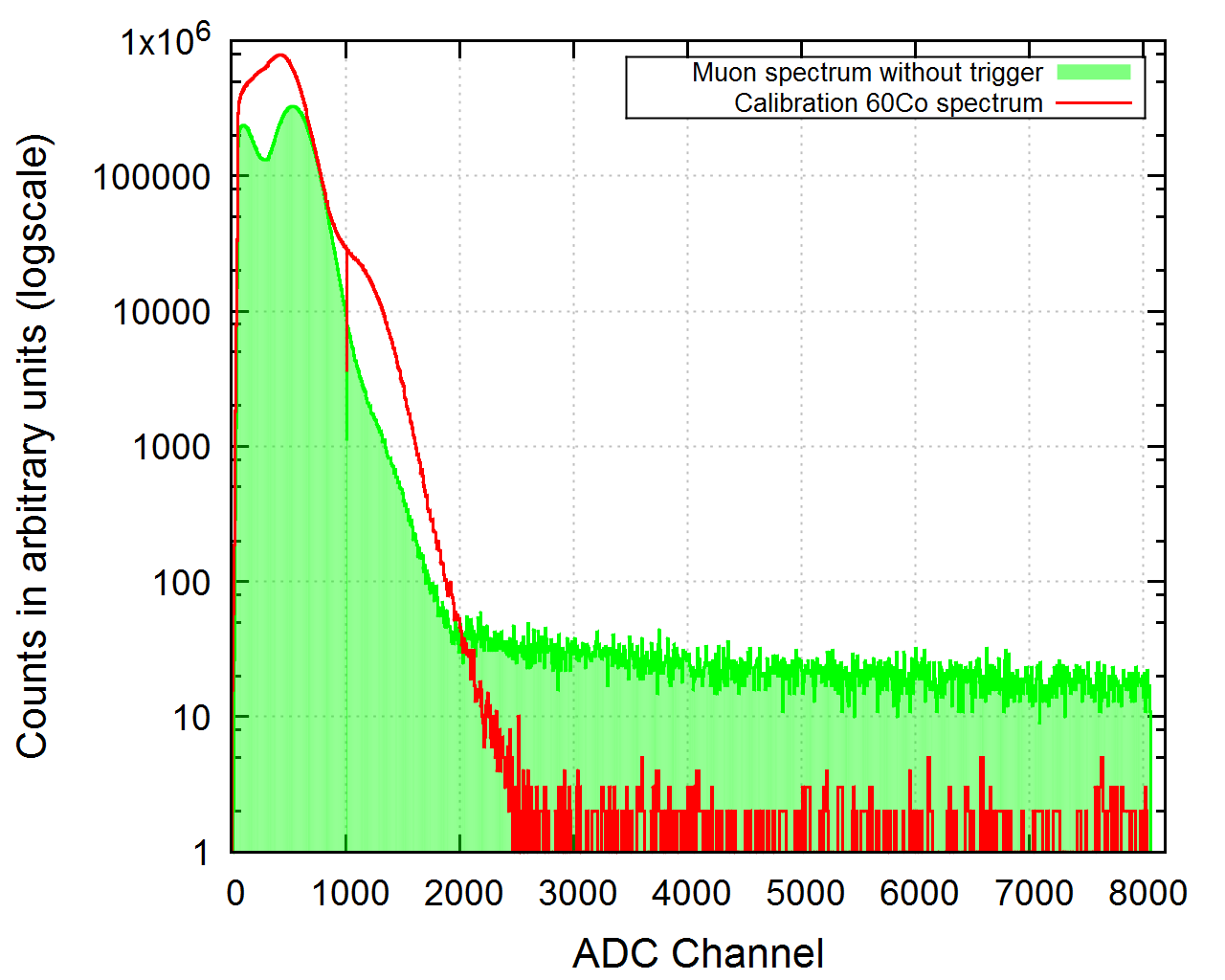}}
	\hfill
	\subfloat[Energy spectrum with upper and lower trigger and wide range] {\includegraphics[width=0.48\textwidth]{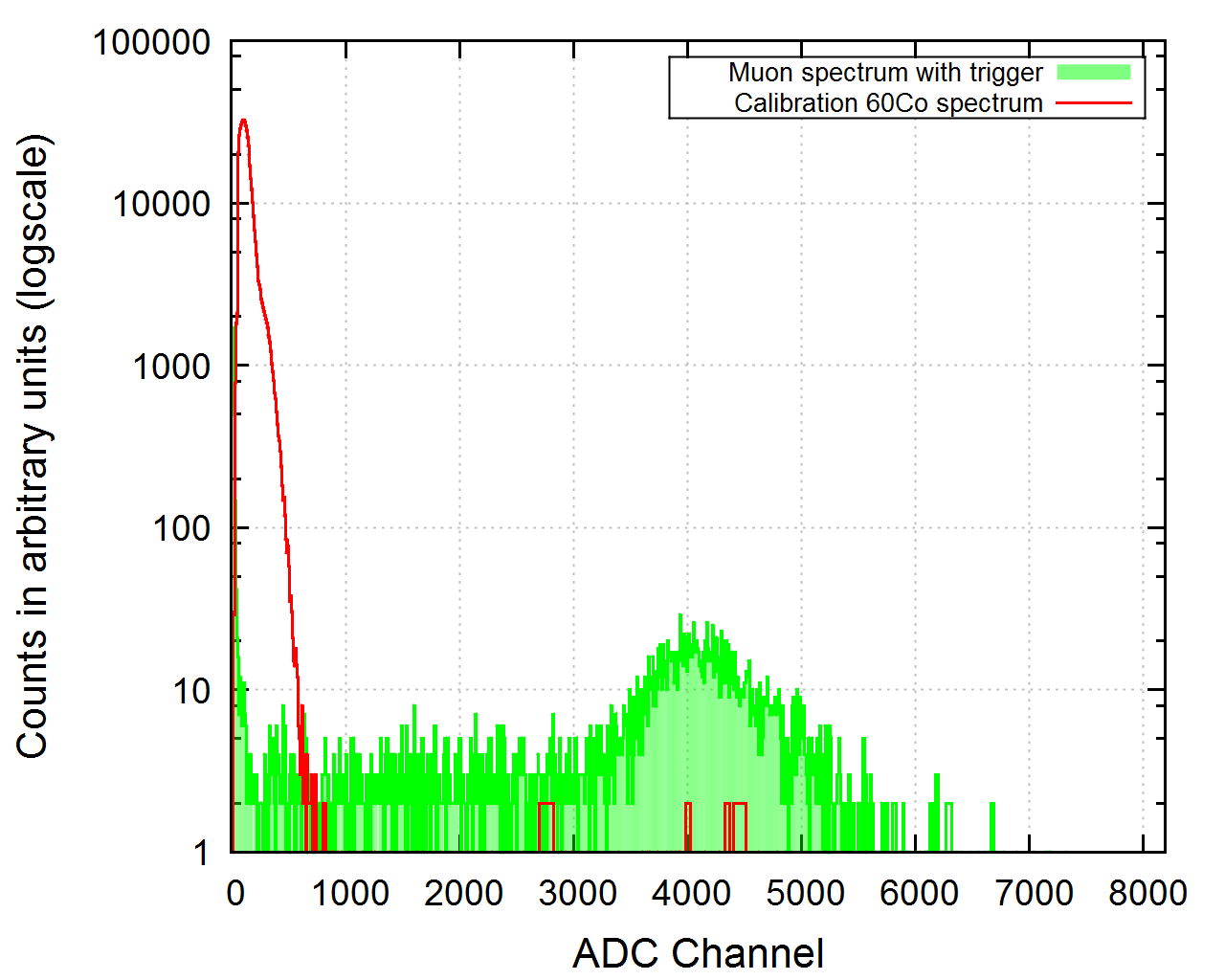}}
	\hfill
	\caption[Energy loss spectrum of cosmic muons]{Energy loss spectrum of cosmic muons. Picture (a) shows a large intrinsic background and a smeared muon spectrum. The chosen dynamic range was chosen too small. Graph (b) shows a measurement with larger dynamic range and as a reference the cobalt spectrum from the calibration run.}
	\label{fig:ch5:muon_spectrum}
\end{figure}

 
Since the light yield is higher and the noise is suppressed at low temperatures, the \pwo{} was cooled down to $\SI{-25}{\degreeCelsius}$. A \co{} spectrum was taken to calibrate the range of the analog-to-digital converter, the spectrum can be seen in figure \ref{fig:ch5:muon_spectrum}. The gain was adjusted to set the energy range of the ADC from $\SI{0}{\MeV}$ to $\SI{40}{MeV}$. After removing the calibration source, data was taken for $\SI{48}{\hour}$ due to the small flux of muons through the crystal. The resulting spectrum is shown in the same figure.  \par 
A large background can be found in the first 2000 channels (see figure \ref{fig:ch5:muon_spectrum}(a)). By comparison with the calibration spectrum of \co{}, it appears that several other gamma sources, either intrinsic or natural environmental radiation, with energies up to $\SI{2}{\MeV}$ contribute to the muon spectrum. A possible candidate is \ka{} with a gamma energy of $\SI{1460.9}{\keV}$, which not only is present in the environment but also as part of the crystal (impurity, up to $0.48$ ppm \cite{analytical_report}). The muon spectrum itself is broad and flat and shows no Landau distribution at all. This might be caused by the geometry of the setup when muons travel through more or less than $\SI{2}{\centi\meter}$ of crystal and hence lose not exactly $\SI{20}{\MeV}$ of energy. This could cause the spectrum to show no distinct Landau peak. \par 
As a solution for this problem, a trigger system  consisting of two plastic scintillator bars with photomultiplier tube readout was installed. Using one bar as an upper and one bar as a lower trigger, the two plastics were placed above and beneath the \pwo{} crystal. The coincident trigger signal gates the ADC, therefore only events measured in all three scintillators were added to the spectrum. The data is plotted in figure \ref{fig:ch5:muon_spectrum}(b). \par 
The first set of data, plot (a), was taken with a smaller energy range and without trigger system. The calibration spectrum was set to roughly $\SI{1200}{\keV}$ at channel 400, using the double peak of \co{}. The trigger system was then introduced since no peak could be observed and due to the large background. Plot (b) shows the result for a broader energy range where the calibration peak could not be used as intended because of the nonlinearity of the ADC in small channel values. A peak at channel 4000 can be observed and and can be more confidently interpreted as a Landau distribution. At lower channels (200 to 3000), smaller energy values due to geometrical effects are visible, for the plastic scintillator bars are larger as the \pwo{} crystal and therefore transition paths smaller than $\SI{2}{\centi\meter}$ are possible for muons.\\ \indent
In this paper we show the outcome of the development of a SIPM-based readout module for organic and inorganic scintillators. For two different configurations, the temperature dependence of the breakdown and operation voltage were examined, as well as the dependency of the dark count rate on discrimination threshold. Timing measurements with the ``hybrid" configurations and energy measurements with the ``parallel" configurations show promising results and indicate that in future applications SiPMs might be a possible substitute for e.g. PMTs or APDs, though some performance issues have been shown up which necessitate further investigation.
\subsubsection*{Acknowledgments}     
This paper was written to summarize findings of the author's bachelor thesis \cite{Lukas_Thesis}. The research and writing of this thesis was performed in Summer 2017 at the Justus Liebig University of Giessen, Germany, under supervision of Dr. Hans-Georg Zaunick and Prof. Dr. Kai-Thomas Brinkmann.  

\bibliography{./bib}

\end{document}